\documentclass[aps,prl,preprint,tightenlines,superscriptaddress,showpacs,floatfix,letterpaper,longbibliography]{revtex4-1}

\usepackage{amsmath,amssymb,amsthm}
\usepackage{color,bm,graphicx,url}

\newcommand\tD{{\bm{\mathsf{D}}}}

\newcommand\vI{{\bm{I}}}
\newcommand\vO{{\bm{O}}}

\newcommand\prob{{\mathbb{P}}}

\newcommand\bv{{\bm{v}}}
\newcommand\bw{{\bm{w}}}

\newcommand\refSI{{(SM Text)}}
\newcommand\refSInb{{SM Text}}

\begin{document}

\title{Information transmission and signal permutation\\in active flow networks}

\author{Francis G. Woodhouse}
\email[]{f.g.woodhouse@damtp.cam.ac.uk}
\affiliation{Department of Applied Mathematics and Theoretical Physics, Centre for Mathematical Sciences, University of Cambridge, Wilberforce Road, Cambridge CB3 0WA, U.K.}

\author{Joanna B. Fawcett}
\altaffiliation{Present address: Department of Mathematics, Imperial College London, London SW7 2AZ, U.K.}
\affiliation{Department of Pure Mathematics and Mathematical Statistics, Centre for Mathematical Sciences, University of Cambridge, Wilberforce Road, Cambridge CB3 0WB, U.K.}

\author{J\"orn Dunkel}
\affiliation{Department of Mathematics, Massachusetts Institute of Technology, 77 Massachusetts Avenue, Cambridge MA 02139-4307, U.S.A.}

\date{\today}

\begin{abstract}
\noindent
Recent experiments show that both natural and artificial microswimmers in narrow channel-like geometries will self-organise to form steady, directed flows. This suggests that networks of flowing active matter could function as novel autonomous microfluidic devices. However, little is known about how information propagates through these far-from-equilibrium systems. Through a mathematical analogy with spin-ice vertex models, we investigate here the input-output characteristics of generic incompressible active flow networks (AFNs). Our analysis shows that information transport  through an AFN is inherently different from conventional pressure or voltage driven networks. Active flows on hexagonal arrays preserve input information over longer distances than their passive counterparts and are highly sensitive to bulk topological defects, whose presence can be inferred from marginal input-output distributions alone. This sensitivity further allows controlled permutations on parallel inputs, revealing an unexpected link between active matter and group theory that can guide new microfluidic mixing strategies facilitated by active matter and aid the design of generic autonomous information transport networks.
\end{abstract}

\maketitle

\section*{Introduction}

Group theory~\cite{2011Artin_Algebra,Rotman} forms the mathematical foundation of ancient~\cite{1996Kahn_Book} and modern~\cite{2014KatzLindell} cryptography. Systematic permutations of the symbols in a given alphabet define the most basic algorithms for encoding information~\cite{1996Kahn_Book,2014KatzLindell}. The efficiency and robustness of such encryption schemes is tightly linked to the structural properties of the underlying permutation groups. This profound connection was first realised by the Polish mathematician Marian Rejewski~\cite{1981Rejewski} in 1932 and, a few years later, used by Alan Turing to decipher codes produced by the Enigma machine~\cite{Hodges_Turing}, a mechanical encoding device employed by the German Army during WWII. Nowadays, information transfer and encryption assume  ever-increasing importance in the development of new technologies, from the internet~\cite{Atzori20102787} and smart phones~\cite{Mayer17052016} to quantum communication~\cite{Hughes1584}. Yet, information transport is also a salient feature of many, if not all, biological systems~\cite{1986Hopfield_Science,2012Bialek,Adamatzky}. This raises interesting conceptual and practical questions as to whether one can use biological or engineered active matter components~\cite{2015Pearce,2016Nicolau_PNAS,2017Woodhouse_NComms} to transport and encrypt information, and how efficiently such active information transportation devices can operate relative to conventional passive information flow networks~\cite{1928Hartley,1949Shannon}.

Here, we explore these questions theoretically by focusing on quasi-incompressible active flow networks (AFNs) that can be realised with dense suspensions of bacteria~\cite{2016Wioland_NPhys,2016Wioland_NJP} or other types of natural or engineered microswimmers~\cite{2008Walther_SM,2012Sanchez_Nature,Bricard2013_Nature,2017Wueaal}.  In contrast to voltage-driven electric~\cite{1949Shannon} or pressure-driven microfluidic~\cite{Prakash832,Fuerstman828} circuits, material and information transport in AFNs is facilitated by the conversion of chemical energy into kinetic energy~\cite{2013Marchetti_Review,2016Wioland_NJP,2013Buttinoni_PRL} at the level of the microscopic constituents, such as bacteria~\cite{2016Wioland_NJP} or Janus particles~\cite{2008Walther_SM}, which can carry information individually or collectively. Building on a mathematical correspondence with discrete spin-ice vertex models~\cite{2016Woodhouse_PNAS}, we will investigate the similarities and differences between the propagation of input signals through internally driven active and externally driven passive flow networks for different lattice geometries. This analysis shows that topological constraints intrinsic to incompressible AFNs enable more robust information flow than in comparable passive networks. In the second part, we will demonstrate how bulk topological defects in AFN lattices can be detected holographically from input--output correlations---that is, from boundary flows alone without any observation of the bulk---and can be utilised to realise specific permutation groups. In doing so, we will establish a fundamental connection between active matter flows in complex topologies and the Cayley graph structure of permutation groups. We conclude by showing how these ideas can be extended to general random graphs to achieve more efficient signal coding.

\vspace{0.5cm}
\section*{Results and Discussion}

\subsection*{Active matter vertex models}

Active matter systems self-organise and spontaneously flow by persistent conversion of chemical energy to stress and are therefore, by nature, non-equilibrium systems.
Their great diversity, encompassing motile cells~\cite{2016Wioland_NPhys,Creppy2016_Interface}, driven microfilaments~\cite{2012Sanchez_Nature} and artificial microswimmers~\cite{2008Walther_SM,Bricard2013_Nature} to name just three classes, means that a wide range of precise behaviours exist. Generically these systems possess non-Boltzmann steady state distributions and non-zero probability currents in state space~\cite{2016Battle_Science}, but certain reductions and limits such as those for coloured noise~\cite{1987Jung_PRA,2016Fodor_PRL} can reveal pseudo-equilibrium behaviour.
In particular, previous work~\cite{2016Wioland_NPhys} has shown that a dense suspension of bacteria confined within a lattice of interacting circular cavities can be captured by a pseudo-equilibrium model in coarse-grained degrees of freedom, namely the average `spin' of each circular cavity. Linear confinement has also been shown experimentally to cause near-unidirectional flow in various active systems~\cite{2016Wioland_NJP,Bricard2013_Nature,Creppy2016_Interface,2017Wueaal}, reducing complex behaviour to a single degree of freedom.
These ideas naturally extend to network-like environments, suggesting that the behaviour of an active flow network (AFN)---that is, a network of narrow channels filled with dense active matter---can be reduced to the coarse-grained mean flow along each channel of the network, represented by directed flows along the edges of a graph.
A dynamical model with active friction leads to slime-mould-like oscillatory pumping states~\cite{2017Forrow_PRL}, while in the dense incompressible limit, a pseudo-equilibrium model based on that verified for circular confinement~\cite{2016Wioland_NPhys} displays topologically-determined stochastic selection of network flow loops~\cite{2016Woodhouse_PNAS} mediated through flow interactions at the mass-conserving junctions between edges. Extending this incompressible model to add inputs and outputs as boundary vertices whose mass flux is controlled or free, respectively, leads to the ability to perform elementary logical operations by appropriate network design~\cite{2017Woodhouse_NComms}.
It is this premise which we adopt here, formalised as follows.

Let $\Gamma$ be a graph with edge set $E$ and vertex set $V \cup \partial\Gamma$, where $V$ is the set of interior vertices and $\partial\Gamma$ are \mbox{degree-1} boundary vertices used as inputs and outputs. Every edge $e \in E$ is assigned an arbitrary orientation, from which we define the $|V|\times |E|$ incidence matrix ${\tD = [D_{ve}]}$ where $D_{ve}$ is $-1$ if edge~$e$ is oriented outwards from vertex~$v$, $+1$ if $e$ points into $v$, and $0$ if $v$ and $e$ are not incident.
A flow configuration $\Phi = (\phi_e)$ on $\Gamma$ is then a vector of signed flows $\phi_e \in \{-1,0,+1\}$ along each $e \in E$, where $\phi_e = +1$ represents flow with the orientation of~$e$ and $\phi_e = -1$ is flow against the orientation of~$e$, so that the flux into vertex~$v$ along edge $e$ is $f_{ve} = D_{ve}\phi_e$.
A non-zero flow $|\phi_e| = 1$ indicates self-organised unidirectional flow along $e$ at the typical velocity of the active matter system under consideration, normalised to unity, while $\phi_e = 0$ corresponds to a quiescent, overturning or turbulent state within the channel with zero net flux. This discretisation of flow states is a simplification of velocities fluctuating within a double-welled potential~\cite{2016Woodhouse_PNAS,2015Chen_NJP}, modelling the tendency of active suspensions to adopt either a unidirectional flow state at a preferred velocity or, failing that, a qualitatively different state~\cite{2016Wioland_NJP}.

The space of permissible flows $\Phi$ is constrained by flux conservation, through which we implement inputs and outputs.
Every internal vertex $v\in V$ must have as many in-flows as out-flows, corresponding to the flux incompressibility condition
\begin{align*}
\sum_e f_{ve} = (\tD \cdot \Phi)_v = 0.
\end{align*}
Inputs and outputs are set and read through the flux at the boundary vertices $\partial\Gamma = \partial\Gamma_\text{in} \cup \partial\Gamma_\text{out}$, where the input vertices $\partial\Gamma_\text{in}$ and output vertices $\partial\Gamma_\text{out}$ are disjoint.
For a given digital input vector $\vI = (I_v) \in \{0,1\}^{|\partial\Gamma_\text{in}|}$ we impose that the vertex $v \in \partial\Gamma_\text{in}$ corresponding to input $I_v$ have net flux
\begin{align*}
\sum_e f_{ve} = (\tD \cdot \Phi)_v = -I_v,
\end{align*}
so that an activated input injects matter into the network.
Output vertices, on the other hand, are left unconstrained to allow matter to flow out of them or not as network interactions dictate; the output vector $\vO = (O_v) \in \{0,1\}^{|\partial\Gamma_\text{out}|}$ for flow state $\Phi$ is then read off as the flow
\begin{align*}
O_v \equiv \sum_e f_{ve} = (\tD \cdot \Phi)_v
\end{align*}
through each $v \in \partial\Gamma_\text{out}$.
Finally, to prevent spurious matter inflow through the outputs, we impose that each edge $e$ incident to an output vertex (of which there is one per output, as outputs have degree 1) only permit flow toward the output, $\phi_e \in \{0,+1\}$. In microfluidic realisations, such an active matter diode can be realised through geometric channel patterning~\cite{Denissenko2012_PNAS}.

To model the spontaneous self-organised flow typical of active matter~\cite{2016Wioland_NPhys,2016Wioland_NJP,2017Wueaal,Creppy2016_Interface}, we adopt a pseudo-equilibrium approach. Define the energy of a configuration $\Phi$ to be
\begin{align*}
H(\Phi) = -\frac{1}{12}\lambda \sum_{e \in E} |\phi_e|,
\end{align*}
with polarisation strength constant $\lambda$ (where the factor of $1/12$ is for consistency with previous continuum models~\cite{2016Woodhouse_PNAS}).
For a fixed input vector $\vI$ we then assume a pseudo-equilibrium model selecting states according to the Boltzmann distribution $p(\Phi | \vI) \propto e^{-\beta H(\Phi)}$ subject to the incompressibility and input--output constraints.
This favours configurations with more flowing edges, as we might expect from active matter systems in confinement~\cite{2016Wioland_NJP,Bricard2013_Nature,Furthauer2012_NJP,Creppy2016_Interface}.
Indeed, the established Toner--Tu model of self-organised flow~\cite{1995Toner_PRL,1998Toner_PRE,2015Chen_NJP} reduces to that of overdamped diffusion in a double-welled potential when averaged along a narrow channel~\cite{2017Woodhouse_NComms}, yielding Boltzmann statistics as per a Landau theory; even if real-world AFNs do not obey exact equilibrium statistics in coarse-grained variables, as is likely, the intrinsic propensity of active matter toward flowing states at characteristic velocities at the heart of the Toner--Tu model suggests that we should expect statistics at least similar to the pseudo-energy fluctuations encoded in the Boltzmann distribution.
The result is a form of vertex model on general graphs in the same family as ice-type or loop models~\cite{Baxter,2011Morgan_NatPhys,1974Wu_JMP,2013Chern_PRE}, endowed with input--output capability, which qualitatively replicates the full continuous lattice field model of Ref.~\cite{2016Woodhouse_PNAS}  \refSI.

\begin{figure*}
\includegraphics[width=\textwidth]{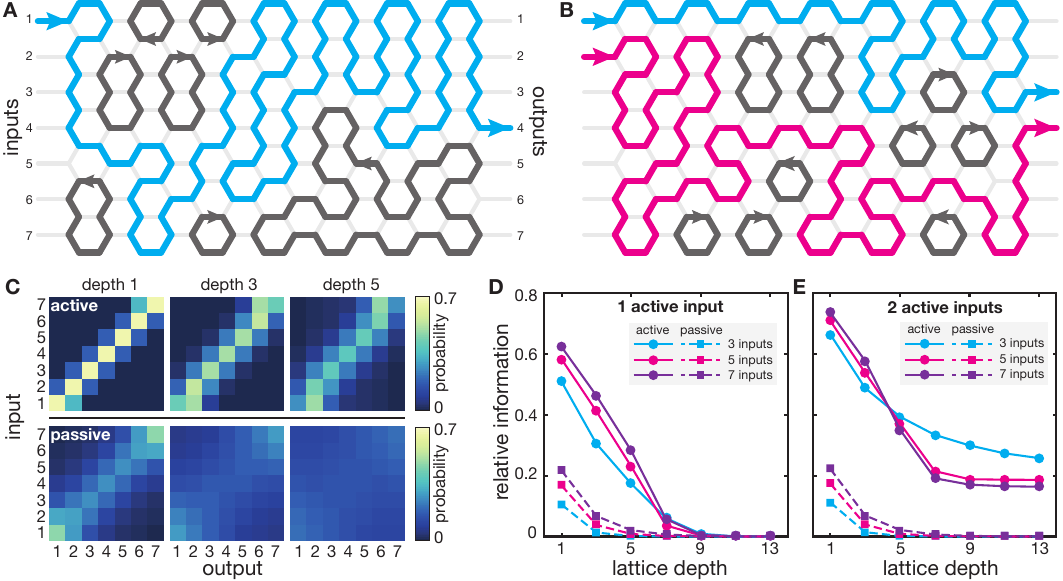}
\caption{
Topological protection of input--output correlations in planar AFNs.
(A) Example configuration of active flow on a $7$-input $13$-deep hexagonal lattice with input $1$ activated, with the input--output flow route highlighted. Thick edges are actively flowing, thin edges are in a zero-flow state.
(B) As in (A) but with inputs $1$ and $2$ both activated. The input--output flows are topologically prohibited from crossing.
(C) Marginal density maps of output distributions with one activated input at low noise $(\beta\lambda)^{-1} = 0.02$. Shown are three depths of hexagonal lattice network for active flow (upper) compared to passive pressure-driven flow (bottom). Each map indicates the probability of an output being activated for a fixed activated input. Active flow data was determined by exhaustive evaluation (depths 1 and 3) and Monte Carlo simulation (depth 5)~\refSI.
(D)~Relative mutual information $U_1(X | Y)$ in determining an activated input $X$, chosen uniformly at random, from observing a particular output~$Y$ at $(\beta\lambda)^{-1} = 0.02$, for one activated input over a range of  hexagonal lattice sizes. Circles denote active flow, squares denote passive flow; comparing the two shows the greater input information retention of active flow across non-trivial networks.
Active flow data computed for small lattices by exhaustive evaluation and other lattices by Monte Carlo simulation~\refSI.
(E) As in (D), but for the information $U_2(X|Y)$ between two activated inputs and outputs. Active flows in planar lattices preserve input ordering, so the mutual information is bounded below by a non-zero constant.}
\label{fig:1}
\end{figure*}

States comprise flowing edges with $|\phi_e| = 1$ and non-flowing edges with $\phi_e = 0$, with flows balanced at every internal vertex and flow out of each activated input.
If we now restrict $\Gamma$ to have vertices of degree at most~$3$, then incompressibility implies that flows become mutually excluding: each internal vertex must have either zero flowing edges or two flowing edges, one in and one out, so stable states comprise non-intersecting flow paths from each activated input to an output with the remaining edges filled by non-intersecting closed cycles of flow (Fig.~\ref{fig:1}A,B).
Since this is where topology design has the greatest potential impact on active flow, we restrict attention to this case here.
We also confine ourselves to the low-noise regime $(\beta\lambda)^{-1} \ll 1$ where appropriate, relevant for strongly confined active matter in a well-controlled environment~\cite{2016Wioland_NJP,2017Wueaal,Wioland2013_PRL}.

\vspace{0.5cm}
\subsection*{Topologically protected information transport}

Active flow networks display markedly different characteristics compared to passive pressure-driven flows or simple random walks.
This is best explored in a lattice topology.
Let $\Gamma$ be an $M \times N$ hexagonal lattice with $N$ inputs and $N$ outputs labelled $i = 1, \ldots, N$, as in Fig.~\ref{fig:1}A.
We refer to the number $M$ of lattice layers between the inputs and the outputs as the depth of $\Gamma$.
When one input is activated,
the AFN picks out a distinct path from the input to an output, with any remaining space filled with vertex-disjoint closed cycles (Fig.~\ref{fig:1}A).
The particular output chosen is probabilistic~\cite{2017Woodhouse_NComms}, and taking an ensemble average (or time average, if dynamics are specified) yields a probability distribution $p(j | i) = \prob(O_j = 1 | I_i = 1)$ for the output from a given input (Fig.~\ref{fig:1}C).

Output from an AFN is markedly different to that of an equivalent passive flow network. We compare AFN output to the steady-state output flux in a linear microfluidic network driven by a fixed constant inflow on $\partial\Gamma_\text{in}$ equal to the input vector~$\vI$, with zero (reference) pressure on $\partial\Gamma_\text{out}$ and equal resistance on every edge (see \refSInb~for the mathematical formulation). Mass conservation implies that this has the same total input--output flux as the active flow network. Upon activation, the edge resistances drive the flow towards a unique attracting steady state, whose distribution of output flux on $\partial\Gamma_\text{out}$ can then be compared to the ensemble-averaged output $\langle \vO \rangle$ of the equivalent AFN. In addition, the steady-state output distribution of this microfluidic network is equivalent to the steady-state probability distribution of a symmetric random walk starting at input $i$ with sinks at the outputs~\cite{Redner}; we henceforth refer to either of these mathematically identical systems as `passive flow'.
With a single input active, so that $I_i = 1$ for the active input $i$ and $I_j = 0$ for $j \neq i$, a passive flow network disperses the input among all output nodes, while the equivalent ensemble-averaged output distribution $\langle \vO \rangle$ from the AFN instead  retains a distinct signature of its input for larger lattices where passive output is near-uniform (Fig.~\ref{fig:1}C).
Thus, the globally exploratory nature of active network flow allows for output from a non-trivial active network to be traceable to the original input, whereas passive flow is virtually untraceable on all lattices.

When multiple inputs are activated, the vertex-disjoint input--output paths mutually exclude one another in AFNs. This alters the output distribution in a fashion dependent on the graph topology, offering additional control over signal propagation compared with passive networks.
Furthermore, the discrete nature of active flow means that each input can be traced to its output without visualising the intervening network by distinctly marking the input flows.
For a planar network such as that in Fig.~\ref{fig:1}, the order of the outputs must match the order of the inputs: if input~$1$ is activated and connects to output~$3$, say, then input~$2$ can only connect to output~$4$ onwards (Fig.~\ref{fig:1}B), in stark contrast to the linearity of passive network flow.
This suggests that active flows may be particularly adept at retaining input configuration memory when more than one input is activated.

The extent to which inputs can be inferred from outputs is captured by the mutual information~\cite{1999Borst_NNeuro}.
Suppose we uniformly at random choose one input $X$ to activate. This connects to one output $Y$ according to a topology-dependent probability distribution $p(y|x)$ (Fig.~\ref{fig:1}C). If we can only measure the output and do not know which input was activated, then how well the activated input can be inferred from an observed output is described by the $1$-input  relative mutual information~\refSI
\begin{align*}
 U_1(X|Y) = \frac{1}{N \log N} \sum_{x,y} p(y|x) \log \left[ \frac{p(y|x)}{p(y)} \right].
\end{align*}
This measures the information gained relative to the maximum possible $\log_2 N$ bits, so that $U_1 = 1$ means exact input--output matching and $U_1 = 0$ means input and output are independent.
The equivalent notion of output observation in the case of passive flow is that of seeing a single random walker arrive at an output for random walks, or observing the destination of a single input tracer particle in pressure-driven microfluidic flow.
Numerically evaluating $U_1$ over a range of hexagonal lattice sizes shows that AFNs preserve input information over notably larger graphs than passive flow (Fig.~\ref{fig:1}D), allowing the activated input to be inferred with high confidence using comparatively few system samples.

With two labelled inputs activated, mutual information~\cite{1999Borst_NNeuro} captures a fundamental difference between AFNs and classical flow.
The randomly chosen activated inputs $X_1$ and $X_2$ are now represented by an ordered pair $X = (X_1,X_2)$ with $X_1 \neq X_2$, where $X_1$ is labelled red and $X_2$ is labelled blue, say.
This yields an output pair $Y = (Y_1, Y_2)$, where $Y_1$ is the output observed red and $Y_2$ the output observed blue, again sampled from a distribution $p(y | x)$. The two-input relative mutual information is then~\refSI
\begin{align*}
 &U_2(X|Y) =
 \frac{1}{N(N-1) \log N(N-1)} \sum_{x,y} p(y|x) \log \left[ \frac{p(y|x)}{p(y)} \right],
\end{align*}
where the modified prefactor reflects the $N(N-1)$ possible labelled input pairs.
Evaluating $U_2$ for AFNs and classical flow (Fig.~\ref{fig:1}E) now yields a qualitative distinction: not only do AFNs preserve information better on larger graphs, as with one input, but $U_2$ asymptotes to a non-zero constant $1 / \log_2 N(N-1)$. This is because mutual exclusion of input--output streams in planar networks means that these AFNs preserve the ordering of their inputs, implying a guaranteed bit of information for even the largest planar lattices.

With regard to applications, the partial topological protection of input--output correlations in planar AFNs suggests interesting possibilities for tuning and enhancing information propagation through the inclusion of auxiliary control currents. Moreover,  as we shall show next, it also allows holographic detection of non-planar lattice defects from input and output distributions alone.

\begin{figure}
\includegraphics{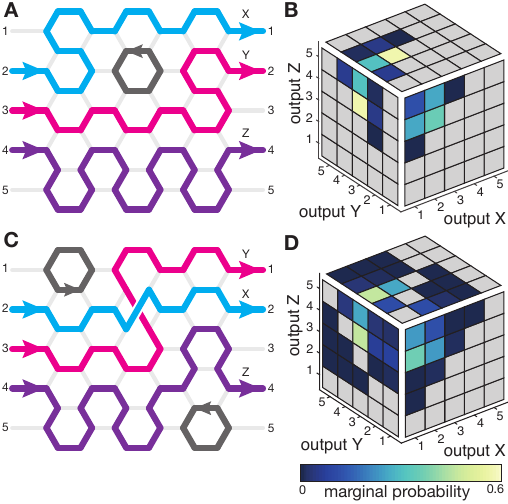}
\caption{
Holographic detection of lattice defects.
(A)~Example active flow on a $5 \times 5$ lattice with 3 activated inputs.
(B)~Joint distribution $p(x,y,z)$ of activated outputs $2 \rightarrow X$, $3 \rightarrow Y$, $4 \rightarrow Z$ for the three activated inputs $2, 3, 4$ in (A) at low noise $(\beta\lambda)^{-1} = 0.02$, shown by the three marginal densities derived from summing over one each of $X$, $Y$ and $Z$ as determined by exhaustive evaluation~\refSI. Grey cells indicate topologically prohibited output orderings violating $X < Y < Z$.
(C)~As in (A) but for a lattice with a planarity defect, allowing input--output streams to cross.
(D)~Output densities as in (B) but now for the defective lattice in (C), demonstrating non-zero probabilities in regions of the distribution previously prohibited by topology.}
\label{fig:2}
\end{figure}

\vspace{0.5cm}
\subsection*{Holographic defect detection}

If $\Gamma$ is not planar then input streams can cross, yielding qualitative changes in the joint distribution of output probabilities compared to that of a similar planar graph.
Suppose, for instance, that inputs $2$, $3$ and $4$ are activated on the $5 \times 5$ hexagonal lattice of Fig.~\ref{fig:2}A. Denoting the activated inputs' respective random outputs by $X$, $Y$ and $Z$, planarity of the lattice means that we must always have $X < Y < Z$ (under our labelling of inputs and outputs as in Fig.~\ref{fig:2}A).
This implies that the joint distribution $p(x,y,z) = \prob(2 \rightarrow x , 3 \rightarrow y , 4 \rightarrow z)$ is only non-zero in the small subspace $x < y < z$ (Fig.~\ref{fig:2}B).
Now, introduce a small planarity defect into the lattice by exchanging endpoints between two horizontal edges of one hexagon (Fig.~\ref{fig:2}C), akin to the rewiring construction of Watts--Strogatz networks~\cite{1998Watts_Nature}.
Two input--output streams can now cross once, allowing output ordering to change and thereby introducing non-zero probabilities within previously prohibited regions of $p(x,y,z)$ (Fig.~\ref{fig:2}D).
This reflection of bulk lattice structure in the surface marginals presents a planarity rejection test if the intervening graph is unknown or difficult to embed.

\vspace{0.5cm}
\subsection*{Realising permutation groups with active flows}

Activating all inputs of a hexagonal lattice with crossover defects results in a stochastic permutation device.
In this case, since there are as many outputs as inputs, each permissible flow configuration defines a bijection $f : \{1,\ldots,N\} \rightarrow \{1,\ldots,N\}$ (that is, a permutation of the integers $1$ to $N$) where input $i$ connects to output $f(i)$.
If the stream at input $i$ is then given label $v_i$, this arrives at output $f(i)$. Denoting the vector of all input labels by  $\bv = (v_i)$ and the vector of output labels by $\bw = (w_j)$, where $w_j$ is the label read at output $j$, any such permutation can be compactly written as~$\bw = \sigma \bv$ for some unique invertible permutation matrix~$\sigma$~\cite{Rotman,2011Artin_Algebra}.
Now, without any crossover defects, planarity implies that the only possible $\sigma$ is the identity map, since flows cannot swap due to the complete topological protection in this case.
However, introducing crossover defects makes non-trivial~$\sigma$ possible. In general, an output configuration of such an AFN consists of a permutation $\sigma \in \Sigma$ randomly chosen from the set $\Sigma$ of all possible permutations, where $\Sigma$, and each permutation's selection probability, is defined by the placement of interior defects.
Furthermore, when all inputs are activated, the lattice topology implies that permissible flow configurations~$\Phi$ all have the same number of flowing edges and hence the same energy $H(\Phi)$. The Boltzmann distribution $\propto e^{-\beta H(\Phi)}$ is therefore uniform, rendering the flow states and permutation selection probabilities independent of the noise strength~$(\beta\lambda)^{-1}$ for these lattices.

As an example, consider the 1-defect lattice in Fig.~\ref{fig:2}C. When all inputs are activated, this can realise three different permutations $f$, mapping $(1,2,3,4,5)$ to one of $(1,2,3,4,5)$, $(1,3,2,4,5)$ or $(2,1,3,4,5)$. These are, respectively, the identity and the transpositions $(23)$ and $(12)$ in group-theoretic cycle notation~\cite{Rotman}. Thus, this lattice has a set $\Sigma = \{\sigma_1, \sigma_2, \sigma_3\}$ of three possible matrices representing these permutations acting on the input vector $\bv$. The first is the $5 \times 5$ identity matrix $\sigma_1 = I_5$, while the second and third read
\begin{align}
\label{eq:AFN_perms}
 \sigma_2 = \begin{pmatrix}
           1 & 0 & 0 & 0 & 0 \\
           0 & 0 & 1 & 0 & 0 \\
           0 & 1 & 0 & 0 & 0 \\
           0 & 0 & 0 & 1 & 0 \\
           0 & 0 & 0 & 0 & 1
          \end{pmatrix},
\qquad
 \sigma_3 = \begin{pmatrix}
           0 & 1 & 0 & 0 & 0 \\
           1 & 0 & 0 & 0 & 0 \\
           0 & 0 & 1 & 0 & 0 \\
           0 & 0 & 0 & 1 & 0 \\
           0 & 0 & 0 & 0 & 1
          \end{pmatrix}.
\end{align}
These have respective selection probabilities $p_1 = 1/3$, $p_2 = 1/2$ and $p_3 = 1/6$, as computed by exhaustive evaluation~\refSI.

\begin{figure}[b]
\includegraphics{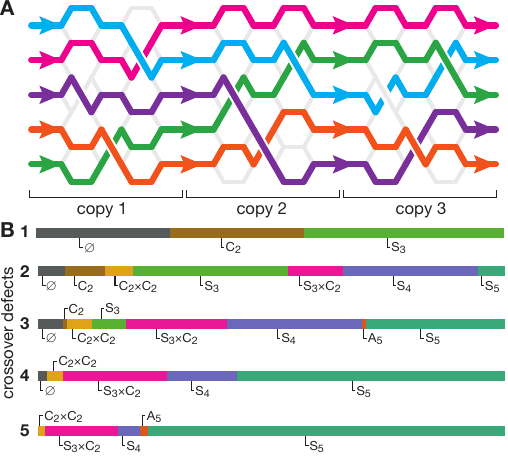}
\caption{Realising permutation groups through AFN concatenation.
(A) Example flow through a 3-fold concatenation of a $3 \times 5$ hexagonal lattice with 3 planarity defects, realising the permutation $(12)(345)$ as the composition $(23)(45) \circ (2354) \circ (12)(45)$.
(B) Groups converged on by repeated concatenation of a $3 \times 5$ lattice with between $1$ to $5$ defects. Group frequencies are shown as their likelihood of occurrence from random defect placements, determined over all possible configurations with each number of defects. The permutation set for each graph was found by exhaustive evaluation from which the group convergence was then evaluated~\refSI. `$\varnothing$' denotes non-convergence in repeated concatenation.}
\label{fig:3}
\end{figure}

In many technological applications, including material mixing and signal encoding, one is interested not in generating a few particular permutations but instead in realising an entire permutation group. For instance, perhaps we wish to employ a microfluidic AFN to combine five components. If we want this to occur in any random order at all, the full symmetric group $S_5$ is called for; alternatively, we may have certain restrictions on ordering---objects $1$ to $3$ must precede objects $4$ and $5$, say---necessitating subgroups of $S_5$ acting on those five objects.
This can be compactly achieved by concatenation of an AFN with copies of itself.
Repeatedly chaining together a small hexagonal lattice containing one or more crossover defects (Fig.~\ref{fig:3}A) causes the input flows to repeatedly permute, akin to a braid~\cite{2015Caussin_PRL,2008Nayak_RMP}, realising different permutation groups according to the lattice defect structure.
Formally, because label permutation $\sigma \bv$ obeys (matrix) composition $(\tau\circ \sigma)\bv  = \tau(\sigma\bv)$, passing the outputs of an AFN $\Gamma$ straight into the inputs of a copy of $\Gamma$ gives a new AFN with permutation set $\Sigma^2 = \{\sigma\tau : \sigma,\tau \in \Sigma\}$
built from all pairwise products of elements in $\Sigma$.
Concatenating a further copy of $\Gamma$ yields an AFN with set $\Sigma^3 = \{\rho \sigma\tau : \rho,\sigma,\tau \in \Sigma\}$, and so-on.
This process either converges, in that there exists an $s$ such that $\Sigma^n = \Sigma^s$ for all $n \geqslant s$, or eventually results in a repeating periodic sequence of permutation sets.
In general, concatenation converges when $\Sigma^k \subseteq \Sigma^{k+1}$ for some~$k$, in which case it must converge on a group~\refSI.
Alternatively, Markov chain theory yields a geometric condition:
concatenation converges precisely when the Cayley graph generated by $\Sigma$ contains a set of cycles whose lengths have a greatest common divisor of~$1$.
The underlying proofs, whose details are given in the \refSInb,  establish a remarkable mathematically rigorous connection between topologically protected active matter flows and the Cayley graph structure of permutation groups, with direct practical implications for material mixing and information encryption.

Continuing the example above, if the network in Fig.~\ref{fig:2}C is concatenated with one copy of itself, the permutations $\Sigma = \{\sigma_1,\sigma_2,\sigma_3\}$ in Eq.~\eqref{eq:AFN_perms} yield a larger set $\Sigma^2 = \{\sigma_1, \sigma_2, \sigma_3, \sigma_4, \sigma_5\}$ of five permutations, where the two new elements read
\begin{align*}
 \sigma_4 = \begin{pmatrix}
           0 & 1 & 0 & 0 & 0 \\
           0 & 0 & 1 & 0 & 0 \\
           1 & 0 & 0 & 0 & 0 \\
           0 & 0 & 0 & 1 & 0 \\
           0 & 0 & 0 & 0 & 1
            \end{pmatrix},
 \qquad
 \sigma_5 = \begin{pmatrix}
           0 & 0 & 1 & 0 & 0 \\
           1 & 0 & 0 & 0 & 0 \\
           0 & 1 & 0 & 0 & 0 \\
           0 & 0 & 0 & 1 & 0 \\
           0 & 0 & 0 & 0 & 1
            \end{pmatrix}.
\end{align*}
These arise as $\sigma_4 = \sigma_2 \sigma_3$ and $\sigma_5 = \sigma_3\sigma_2$, representing the permutations $(132)$ and $(123)$ in cycle notation~\cite{Rotman}.
Concatenating a second copy of the network results in the $6$-element set $\Sigma^3 = \Sigma^2 \cup \{\sigma_6\}$, where $\sigma_6$ represents the transposition $(13)$.
Any further concatenation creates no new permutations---that is, $\Sigma^n = \Sigma^3$ for $n \geqslant 3$---so the concatenation converges, in this case to the symmetric group $S_3$ acting on the first three inputs.
Observe that convergence was guaranteed by finding $\Sigma \subseteq \Sigma^2$, itself a consequence of $\Sigma$ containing the identity~\refSI.

Through concatenation, a variety of groups can be constructed.
Figure~\ref{fig:3}B illustrates the relative abundance of the groups generated by repeated concatenation of $3 \times 5$ lattices with up to $5$ local crossover defects, determined by evaluating all possible networks with each number of defects.
The largest possible group on five inputs, the symmetric group $S_5$, is present, along with six of its 14 non-trivial non-isomorphic proper subgroups.
In fact, $S_5$ can be generated even with only two defects, but this comes at the expense of many concatenations; as the number of defects increases, the probability of generating $S_5$ rises~\cite{2015Morgan_ArchMath} and the requisite number of concatenations falls~(Fig.~S1).
The swapping performed by the local crossover defects is reflected in the subgroups generated: all but $A_5$ are precisely those that can be generated by a set of transpositions~\cite{2005Cameron}.
Notable among the absences are the familiar dihedral groups $D_{8}$ and $D_{10}$, the symmetry groups of the square and pentagon, respectively.
Though $S_3 \times C_2$ is generated frequently and is technically isomorphic to the hexagonal symmetry group $D_{12}$, it only appears here through the natural action of $S_3 \times C_2$ on $5$ points rather than as hexagonal symmetries of $6$ points.
To generate these particular group actions necessitates more complex fundamental permutations than the local swap defects we consider here. In fact, these can be generated by more general AFN topologies, as we will soon describe.

Repeated concatenation continues to have a quantitative effect beyond the point where the qualitative effect ends.
As more copies are added on, even if the permutation set $\Sigma^n$ is constant as $n$ increases, the underlying probability of generating each element of $\Sigma^n$ changes with each additional copy.
Provided $\Sigma^n$ converges to a group, Markov chain theory implies that these probabilities approach the uniform distribution in the limit $n \rightarrow \infty$~\refSI, allowing fine-grained control over output frequencies.
To exemplify this, consider once more the network in Fig.~\ref{fig:2}C and let $G = \Sigma^3$ be the group it generates by concatenation.
If we let $P_{ij}$ be the probability that a single copy of the AFN permutes state $g_i \in G$ to state $g_j \in G$, then the probability that our initial unpermuted (identity) state $g_1$ is sent to $g_i \in G$ after~$n$ concatenations is $P^n_{1i}$.
Using the permutation probabilities found above, this transition matrix reads~\refSI
\begin{align*}
[P_{ij}]
= \begin{pmatrix}
 	1/3 & 1/2 & 1/6 & 0 & 0 & 0 \\
 	1/2 & 1/3 & 0 & 0 & 1/6 & 0 \\
 	1/6 & 0 & 1/3 & 1/2 & 0 & 0 \\
 	0 & 0 & 1/2 & 1/3 & 0 & 1/6 \\
 	0 & 1/6 & 0 & 0 & 1/3 & 1/2 \\
 	0 & 0 & 0 & 1/6 & 1/2 & 1/3 \\
 \end{pmatrix},
\end{align*}
with row and column indexes corresponding to the permutations as before.
Then at $n=3$ we have non-uniform probabilities---$P^3_{12} \approx 0.32$ versus $P^3_{16} \approx 0.06$, for example---but by $n=20$ these have converged to $1/6$ at two decimal places.

\begin{figure}[b]
\includegraphics{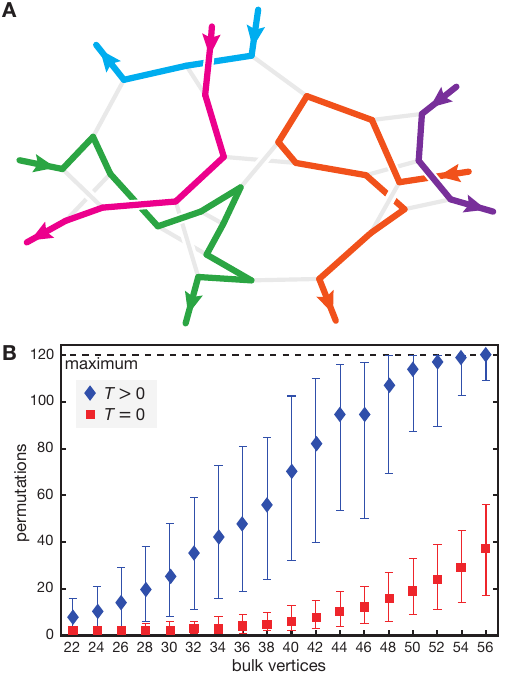}
\caption{
Efficient generation of permutations in random AFNs.
(A)~Example of active flow on a random cubic network with 5 inputs and outputs.
(B)~Number of distinct permutations realised by a sample of random $5$-input/output cubic AFNs at fixed numbers of bulk vertices, at both non-zero (blue diamonds) and zero (red squares) noise strength $T = \beta^{-1}$.
Graphs were generated by 10 random choices of 5 input and output insertions in each of 50 samples of non-isomorphic random cubic graphs, discarding networks possessing no valid ground state when all inputs are activated~\refSI.
Markers denote sample median, bars are 10\% and 90\% quantiles.}
\label{fig:4}
\end{figure}

\vspace{0.5cm}
\subsection*{General random graphs}
Finally, we turn to permutations realised by general graphs.
As the internal structure linking inputs to outputs becomes topologically more complex, many more permutations can often be realised with the same number of internal vertices than in a defect-riddled hexagonal lattice, trading complexity for functionality.
General graphs are also typically not independent of $\beta$, commonly realising far more permutations when $T = \beta^{-1} > 0$ beyond the ground-state permutations seen when $T = 0$.
To illustrate this, we computed the exact number of permutations~$|\Sigma|$ output by samples of input--output-augmented random cubic graphs at fixed (necessarily even) numbers~$V$ of bulk vertices for~$V$ between 22 and 56~\refSI.
As shown in Fig.~\ref{fig:4}, $|\Sigma|$ clearly tends to increase with~$V$ in both the $T = 0$ and $T > 0$ regimes. Indeed, the majority of 56-vertex graphs attain all possible $5! = 120$ permutations when $T > 0$, thus directly outputting $S_5$ with fewer vertices than any of the $S_5$-generating concatenations in Fig.~\ref{fig:3}.

When repeatedly concatenated, zero-noise random AFNs generate a large zoo of $S_5$ subgroups. Upon analysing the random sample in Fig.~\ref{fig:3}B, we found $T=0$ AFNs generating almost all subgroups of $S_5$, including the familiar $D_8$ and $D_{10}$ missing from the hexagonal lattices of Fig.~\ref{fig:2}B.
This suggests that general AFNs at $T=0$ can realise almost any desired group action on repeated concatenation. A less exotic list of groups is generated when $T>0$, identical to those in Fig.~\ref{fig:2}B. However, the elements of rarer groups can likely still be realised with high probability provided $T$ is low and the number of concatenations is as few as possible.

\vspace{0.5cm}
\section*{Conclusion}

To conclude, recent technological advances in the fabrication of soft~\cite{2017Liu_PNAS,2017Lie} and fluid-based~\cite{2008Walther_SM,Bricard2013_Nature,2011Shashi,2013Marchetti_Review}  active materials demand
novel theoretical and algorithmic ideas to guide the functional design of autonomous logical units~\cite{2015Pearce,2017Woodhouse_NComms,2016Nicolau_PNAS},
pattern recognition systems~\cite{2016Fang} and information transport devices operating far from thermal equilibrium.
Vertex models that account for the relevant physical conservation laws and locally driven matter fluxes offer a flexible
testbed for exploring generic properties and limitations of signal transduction in active systems. Building on this framework,
our analysis shows how topological constraints inherent to quasi-incompressible  AFNs can be utilised to
realise the actions of fundamental symmetry groups underlying discrete mixing processes and standard signal encryption protocols,
providing a conceptual basis for potential future implementation of such processes using active matter-based devices.

The planar and non-planar network designs proposed and investigated here could be implemented and tested in microfluidic chips, exploiting recent progress in 3D printing~\cite{2016Bhattacharjee_LapChip} and in the geometric control of collective transport in dense suspensions of microorganisms~\cite{2016Wioland_NJP,2016Wioland_NPhys} and ATP-powered microtubule bundles~\cite{2017Wueaal}.
Furthermore, recent progress in experimental realisation of artificial magnetic and colloidal spin-ice systems~\cite{2011Morgan_NatPhys,2016OrtizAmbriz_NComms,2016Loehr_PRL} suggests that the input--output spin-ice model studied here could itself be directly realised.
 More broadly, however, the above results establish a direct link between active matter and ostensibly unrelated mathematical concepts in information and group theory, thus promising novel symmetry-based approaches to autonomous network design.

\section*{Acknowledgments}
This work was supported by Trinity College, Cambridge (F.G.W.), the London Mathematical Society (J.B.F.), an Alfred P. Sloan
Research Fellowship (J.D.), an Edmund F. Kelly Research Award (J.D.), NSF Award CBET-1510768 (J.D.) and a Complex
Systems Scholar Award of the James S. McDonnell Foundation (J.D.).

\end{document}